\documentclass[preprint,prc,aps,showpacs,showkeys,groupedaddress,floatfix,superscriptaddress]{revtex4-1}
\usepackage{epsfig}
\usepackage{dcolumn}
\usepackage{bm}
\usepackage[utf8]{inputenc}
\usepackage{graphics}
\usepackage{graphicx}
\usepackage{epsfig}
\usepackage{amssymb}
\usepackage{amsmath}
\usepackage{color}
\usepackage{bbold}
\usepackage{subcaption}

\begin{document}

\title{Klein–Gordon Oscillator with Scalar and Vector Potentials in Topologically Charged Ellis–Bronnikov-type Wormhole }

\author{Abbad Moussa}
\email{moussa.abbad@univ-tebessa.dz}
\affiliation{Laboratory of Applied and Theoretical Physics, Chahid Cheikh Larbi-Tébessi University. Tebessa, Algeria}

\author{Houcine Aounallah}
\email{houcine.aounallah@univ-tebessa.dz}
\affiliation{Department of Science and Technology. Chahid Cheikh Larbi-Tébessi University. Tebessa, Algeria}

\author{Prabir Rudra}
\email{prudra.math@gmail.com}
\affiliation{Department of Mathematics, Asutosh College, Kolkata-700026, India}

\author{Faizuddin Ahmed}
\email{faizuddinahmed15@gmail.com}
\affiliation{Department of Physics, University of Science \& Technology Meghalaya, Ri-Bhoi, Meghalaya-793101, India}

\date{\today}
\begin{abstract}
In this work, we study the Klein-Gordon oscillator with equal scalar and vector potentials in a topologically charged  Ellis-Bronnikov wormhole space-time background. The behaviour of a relativistic oscillator field is studied with a position dependent mass via transformation $M^2 \to (M+S(x))^2$ and vector potential through a minimal substitution in the wave equation. Simplifying the Klein-Gordon oscillator equation for three different types of potential, such as linear confining, Coulomb-type, and Cornell-type potential and we arrive a second order differential equation known as the biconfluent Heun (BCH) equation and the corresponding confluent Heun function. Finally, we solve the wave equation by the Frobenius method as a power series expansion around the origin and obtain the energy levels and the wave function.  

\end{abstract}
\maketitle 

\section{Introduction}
It is known that phase transitions are associated with breaking of symmetry in condensed matter physics. Analogically it is expected to have objects in cosmological systems which can be termed as topological defects \cite{td1}. From the physical point of view topological defects are structures that divide a physical system into two or more physical states \cite{td2}. However from the mathematical point of view, they are simply solutions of non-linear differential equations. The most familiar topological defects are either of the point type or linear in nature. Cosmic strings \cite{cs1, cs2} are the most common examples of linear topological defects. Other than this both space-like and time-like dislocations \cite{dl1, dl2} are also well-known candidates for the linear topological defects. In the context of extragalactic astrophysics the point-like defect, global monopole (GM) \cite{gm1} is the most widely found candidate in literature. 

A point-like global monopole (PGM) has been extensively studied in the literature \cite{gm2, gm3, gm4, gm5}. PGM has also been studied on the Dirac and Klein-Gordon quantum oscillators \cite{kg1}. Extensive studies of GM can also be found on quantum harmonic oscillator \cite{qho1, qho2, SR2, MM, MM2}. For non-relativistic quantum systems, a point-like global monopole can actually represent vacancies or impurities \cite{imp}, which in turn can affect the permitted energy levels of the system \cite{energy1}. In addition, PGM has been studied on a particle subjected to a self-interaction potential \cite{nr1}, and non-relativistic particles interact with potential under AB-flux field \cite{MP2,MP3,MP4}. 

The Klein-Gordon oscillator (KGO) is perhaps the most popular form of quantum oscillators in relativistic quantum mechanics, because it can be used to specify the corresponding non-relativistic quantum harmonic oscillator described by the Schrodinger's equation \cite{sch1}. Initially inspired by the Dirac oscillator \cite{dirac1} KGO has recently been investigated under various set-up. Some of the scenarios under which KGO has been studied are Kaluza-Klein theory \cite{kkt}, anti de-Sitter spacetime \cite{ads}, spacetime with torsion \cite{st}, cosmic string spacetime \cite{css}, under the effect of central potentials \cite{cp1, cp2}, etc.

General relativity (GR) admits various classes of solutions out of which wormholes \cite{wh1, wh2} are quite well known and extensively studied. This is because wormholes provide short cuts between different points in spacetime located far away from each other, or even between two different universes. These are hypothetical objects which needs the violation of null energy condition to sustain itself. Wormholes are hypothetical geometric objects having tubular structure that is asymptotically flat on both sides. Wormholes have been investigated from various aspects in literature \cite{wh3, wh4, wh5, wh6, wh7, wh8, wh9, wh10, wh11, wh12, wh13}. One of the simplest and earliest wormhole solution was the Ellis-Bronnikov wormhole \cite{eb1, eb2}. Modified gravity theories \cite{mg1} are considered as an alternative avenue to exotic matter components (dark energy) \cite{de1}. There are numerous candidates among which the Eddington inspired Born-Infield gravity (EiBI) \cite{eibi1} has been studied extensively in literature. This is primarily because of its ability to avoid singularities without resorting to any form of exotic matter. Ellis-Bronnikov wormhole is a solution obtained in the background of EiBI gravity, which describes a static and spherically symmetrical spacetime with the topological charge of GM \cite{eibi2, eibi3}. This solution was obtained by coupling the energy momentum tensor connected with the region exterior to the GM core with the spacetime geometry. This wormhole solution has been studied quite extensively in literature. Scalar field and deflection of light under the effects of topologically
charged Ellis–Bronnikov-type wormhole spacetime was studied in ref. \cite{eb3}. A study on the KGO in topologically charged Ellis–Bronnikov-type wormhole spacetime was performed by Soares et. al. in ref. \cite{eb4}. Motivated by this here we want to extend the study of KGO in topologically charged Ellis–Bronnikov-type wormhole spacetime by including a scalar potential. The paper is organized as follows: In section II we study the KGO under the scalar potential in the Ellis-Bronnikov wormhole spacetime. In the section III we end the paper with some concluding remarks.

\section{Klein–Gordon oscillator with Equal scalar and Vector potential in EB-type Wormhole} 

We want to study the behavior of a relativistic scalar particle with position-dependent mass in a simple Ellis-Bronnikov-type wormhole. Here this position-dependent mass by introducing a scalar potential as a modification in the mass term of the Klein-Gordon equation as the form $m\rightarrow m+S(x)$ where $S(x)$ is the scalar potential, we can be introduce the generalized Klein-Gordon oscillator as the form 
\begin{eqnarray}
\frac{1}{\sqrt{-g}}\left(\partial_{\mu}+m\omega x_{\mu}\right)\left(\sqrt{-g}g^{\mu\nu}\right)\left(\partial_{\nu}-m\omega x_{\nu}\right)\Psi-\left(m+S\left(x\right)\right)^{2}\Psi=0,
\label{kg1}
\end{eqnarray}
where $g=det(g_{\mu\nu})$, $g^{\mu\nu}$ is the inverse metric tensor, m is the rest mass of the scalar field and $x_{\mu}=(0,x,0,0)$. The simple Ellis-Bronnikov-type wormhole is described by the line element $(c=\hbar=1)$
\begin{eqnarray}
ds^{2}=-dt^{2}+\frac{dx^{2}}{\alpha^{2}}+(x^{2}+a^{2})\left(d\theta^{2}+\sin^{2}\theta d\phi^{2}\right),
\label{kg2}
\end{eqnarray}
where $-\infty<x<\infty$, $0<\alpha=1-8\pi^2G\eta_0^2<1$, with $G$ being the universal constant of gravitation, $\alpha$ is the topological defect parameter, and $\eta_0$ the dimensionless volumetric mass density of the point-like GM, and $a=\text{constant}$ is the radius of the wormhole throat. 

In the interaction, we introduce a minimal coupling $\partial_{\mu}\rightarrow\partial_{\mu}+iqA_{\mu}$, where $q$ is the electric charge and $A_{\mu}$ is the electromagnetic four-vector potential $A_{\mu}=\left(-A_{0},\vec{A}\right)$. The Klein-Gordon equation (\ref{kg1}) becomes
\begin{eqnarray}
\frac{1}{\sqrt{-g}}\left(\partial_{\mu}+iqA_{\mu}+m\omega x_{\mu}\right)\left(\sqrt{-g}g^{\mu\nu}\right)\left(\partial_{\nu}+iqA_{\nu}-m\omega x_{\nu}\right)\Psi-\left(m+S\left(x\right)\right)^{2}\Psi=0.
\label{kg3}
\end{eqnarray}

Choosing $\vec{A}=0$ in eqn.(\ref{kg3}) we have,
\begin{eqnarray}
\left(i\frac{\partial}{\partial t}+qA_{0}\right)^{2}\Psi+\alpha^{2}\frac{\partial^{2}\Psi}{\partial x^{2}}+\frac{2\alpha^{2}x}{(x^{2}+a^{2})}\frac{\partial\Psi}{\partial x}-\alpha^{2}m\omega\Psi-\frac{2\alpha^{2}m\omega x^{2}}{(x^{2}+a^{2})}\Psi-\alpha^{2}m^{2}\omega^{2}x^{2}\Psi \nonumber
\end{eqnarray}
\begin{eqnarray}
+\frac{1}{(x^{2}+a^{2})}\left[\frac{1}{\sin\theta}\frac{\partial}{\partial\theta}\left(\sin\theta\frac{\partial\Psi}{\partial\theta}\right)+\frac{1}{\sin^{2}\theta}\frac{\partial^{2}\Psi}{\partial\phi^{2}}\right]-\left(m+S\left(x\right)\right)^{2}\Psi=0.
\label{kg4}
\end{eqnarray}

To solve eqn.(\ref{kg3}) we need to separate it as a set of second order differential equations in terms of each spherical coordinate. Hence we shall consider
\begin{eqnarray}
\Psi\left(t,x,\theta,\phi\right)=R\left(x\right)Y_{l,m}\left(\theta,\phi\right)e^{-iEt}.
\label{kg5}
\end{eqnarray}

By substituting $qA_{0}=V\left(x\right)$ in eqn.(\ref{kg4}) we have
\begin{eqnarray}
\frac{d^{2}R\left(x\right)}{\partial x^{2}}+\frac{2x}{(x^{2}+a^{2})}\frac{dR\left(x\right)}{\partial x}-m\omega R\left(x\right)-\frac{2m\omega x^{2}}{(x^{2}+a^{2})}R\left(x\right)-m^{2}\omega^{2}x^{2}R\left(x\right) \nonumber
\end{eqnarray}
\begin{eqnarray}
+\frac{\left(E+V\left(x\right)\right)^{2}}{\alpha^{2}}R\left(x\right)-\frac{l\left(l+1\right)}{\alpha^{2}(x^{2}+a^{2})}R\left(x\right)-\frac{\left(m+S\left(x\right)\right)^{2}}{\alpha^{2}}R\left(x\right)=0,
\label{kg6}
\end{eqnarray}
where we have used the definition
\begin{eqnarray}
\left[\frac{1}{\sin\theta}\frac{\partial}{\partial\theta}\left(\sin\theta\frac{\partial}{\partial\theta}\right)+\frac{1}{\sin^{2}\theta}\frac{\partial^{2}}{\partial\phi^{2}}\right]Y_{l,m}\left(\theta,\phi\right)=-l\left(l+1\right)Y_{l,m}\left(\theta,\phi\right).
\label{kg7}
\end{eqnarray}
We assume the solution of the second order differential equation (\ref{kg4}) in the form
\begin{eqnarray}
R\left(x\right)=e^{-\frac{1}{2}m\omega x^{2}}Q\left(x\right).
\label{kg8}
\end{eqnarray}
Using this form of solution one can obtain from eqn.(\ref{kg7}) by setting $S(x)=V(x)$ as
\begin{eqnarray}
\frac{d^{2}\,Q(x)}{dx^{2}} +2\,x\,\left[\frac{1}{(x^{2}+a^{2})}-m\,\omega\right]\frac{dQ(x)}{dx}+\Bigg[\frac{2(E-m)S(x)}{\alpha^{2}}+b^2-\frac{4\,m\,\omega x^2+\iota^2}{(x^{2}+a^{2})}\Bigg]\,Q(x)=0,
\label{kg9}
\end{eqnarray}
where we define the new parameters as,
\begin{eqnarray}
b^{2}=\frac{E^{2}-m^{2}-2m\omega\alpha^{2}}{\alpha^{2}}\quad,\quad \iota^{2}=\frac{l\left(l+1\right)}{\alpha^{2}}.
\label{kg10}
\end{eqnarray}

We apply a coordinate transformation $z=-\frac{x^{2}}{a^{2}}$, and rewrite eqn.(\ref{kg9}) as
\begin{eqnarray}
\frac{d^{2}Q\left(z\right)}{dz^{2}}+\left[a^{2}m\omega+\frac{1}{2z}+\frac{1}{\left(z-1\right)}\right]\frac{dQ\left(z\right)}{dz} \nonumber
\end{eqnarray}
\begin{eqnarray}
+\left[\frac{\iota^{2}-a^{2}b^{2}}{4z}+\frac{4a^{2}m\omega-\iota^{2}}{4\left(z-1\right)}+\frac{a^2\,(m-E)\,S(z)}{2\,\alpha^{2}z}\right]Q=0.
\label{kg11}
\end{eqnarray}

Below, we will consider some interaction potentials, such as linear confining potential, Coulomb-type potential and Cornell-type potential and solve the wave equation. 

\subsection{Linear Potential}

In this section, we will study the above quantum system with a linear confining potential $S \propto z$. This type of potential has been used in confinement phenomena in quarks \cite{gg1}, in atomic and molecular physics \cite{gg2,gg6}, in quantum motions of spin-zero and spin-half particles \cite{gg3,gg4,gg5,gg7,gg8,gg9,gg15,gg16}, and in the context of the Kaluza-Klein theory \cite{gg10,gg11,gg12,gg13,gg14}. Throughout the analysis, we have chosen equal scalar vector potentials to study the quantum motions of oscillator field. The linear scalar and vector potentials are given by 

\begin{eqnarray}
V\left(z\right)=S\left(z\right)=C_{1}\,z, \quad C_1>0.
\label{kg12}
\end{eqnarray}

Thereby, substituting potential (\ref{kg12}) in the Eq (\ref{kg11}),  one arrive the following equation
\begin{eqnarray}
\frac{d^{2}Q\left(z\right)}{dz^{2}}+\left[a^{2}m\omega+\frac{1}{2z}+\frac{1}{\left(z-1\right)}\right]\frac{dQ\left(z\right)}{dz} \nonumber
\end{eqnarray}
\begin{eqnarray}
+\left[\frac{\iota^{2}-a^{2}b^{2}}{4z}+\frac{4a^{2}m\omega-\iota^{2}}{4\left(z-1\right)}+\frac{a^{2}\left[m-E\right]C_{1}}{2\alpha^{2}}\right]Q\left(z\right)=0,
\label{kg13}
\end{eqnarray}
where we have set the parameters
\begin{eqnarray}
Q\left(z\right)=e^{\frac{az\left(-am\omega\alpha+\sqrt{2\left(E-m\right)C_{1}+a^{2}m^{2}\omega^{2}\alpha^{2}}\right)}{2\alpha}}\times \nonumber
\end{eqnarray}
\begin{eqnarray}
H_{c}\left(\frac{a\sqrt{2\left(E-m\right)C_{1}+a^{2}m^{2}\omega^{2}\alpha^{2}}}{\alpha},-\frac{1}{2},0,\frac{a^{2}m\omega-a^{2}b^{2}}{4},\frac{a^{2}m\omega+1-\iota^{2}+a^{2}b^{2}}{4};z\right).
\label{kg14}
\end{eqnarray}

Let us consider a power series solution of the function $H (z)$ as follows 
\begin{eqnarray}
H(z)=\sum_{j=0}^{\infty}c_{j}z^{j}.
\label{kg15}
\end{eqnarray}
into the equation (\ref{kg13}), we obtain the recurrence relation
\begin{eqnarray}
c_{j+2}= \nonumber
\end{eqnarray}
\begin{eqnarray}
\frac{\left(2\left(j+1\right)\left(2j+3\right)-\frac{a\sqrt{2\left(E-m\right)C_{1}+a^{2}m^{2}\omega^{2}\alpha^{2}}}{\alpha}\left(4j+5\right)-\iota^{2}+a^{2}b^{2}+a^{2}m\omega\right)}{2\left(j+2\right)\left(2j+3\right)}c_{j+1} \nonumber
\end{eqnarray}
\begin{eqnarray}
+\frac{\left(\frac{a\sqrt{2\left(E-m\right)C_{1}+a^{2}m^{2}\omega^{2}\alpha^{2}}}{\alpha}\left(4j+3\right)+a^{2}m\omega-a^{2}b^{2}\right)}{2\left(j+2\right)\left(2j+3\right)}c_{j}
\label{kg16}
\end{eqnarray}
with the few coefficients
\begin{eqnarray}
c_{1}=\frac{\left(a^{2}b^{2}-\iota^{2}+a^{2}m\omega-\frac{a\sqrt{2\left(E-m\right)C_{1}+a^{2}m^{2}\omega^{2}\alpha^{2}}}{\alpha}\right)}{2}c_{0}.
\label{kg17}
\end{eqnarray}

Let us consider $j=n-1$ when $c_{n+1}=0$ we have
\begin{eqnarray}
c_{n}=\frac{\left(a^{2}b^{2}-\frac{a\sqrt{2\left(E-m\right)C_{1}+a^{2}m^{2}\omega^{2}\alpha^{2}}}{\alpha}\left(4n-1\right)-a^{2}m\omega\right)}{\left(2n\left(2n+1\right)-\frac{a\sqrt{2\left(E-m\right)C_{1}+a^{2}m^{2}\omega^{2}\alpha^{2}}}{\alpha}\left(4n+1\right)-\iota^{2}+a^{2}b^{2}+a^{2}m\omega\right)}c_{n-1}.
\label{kg18}
\end{eqnarray}

For $n=1$, we have 
\begin{eqnarray}
c_{1}=\frac{\left(a^{2}b^{2}-\frac{3a\sqrt{2\left(E-m\right)C_{1}+a^{2}m^{2}\omega^{2}\alpha^{2}}}{\alpha}-a^{2}m\omega\right)}{\left(6-\frac{5a\sqrt{2\left(E-m\right)C_{1}+a^{2}m^{2}\omega^{2}\alpha^{2}}}{\alpha}-\iota^{2}+a^{2}b^{2}+a^{2}m\omega\right)}c_{0}.
\label{kg19}
\end{eqnarray}

By substituting Eq. (\ref{kg19}) into Eq. (\ref{kg17}), we have
\begin{eqnarray}
E_{l,1}^{4}+\left(4\frac{\alpha^{2}}{a^{2}}-2\frac{\alpha^{2}}{a^{2}}\iota^{2}-2m\omega\alpha^{2}-2m^{2}\right)E_{l,1}^{2} \nonumber
\end{eqnarray}
\begin{eqnarray}
-6\frac{\alpha}{a}\sqrt{2\left(E_{l,1}-m\right)C_{1}+a^{2}m^{2}\omega^{2}\alpha^{2}}E_{l,1}^{2} \nonumber
\end{eqnarray}
\begin{eqnarray}
+\left(6\frac{\alpha}{a}m^{2}+\frac{6}{a}m\omega\alpha^{3}+6\iota^{2}\frac{\alpha^{3}}{a^{3}}\right)\sqrt{2\left(E_{l,1}-m\right)C_{1}+a^{2}m^{2}\omega^{2}\alpha^{2}} \nonumber
\end{eqnarray}
\begin{eqnarray}
+m^{4}+2m^{3}\omega\alpha^{2}-4\frac{\alpha^{2}}{a^{2}}m^{2}+2\frac{\alpha^{2}}{a^{2}}\iota^{2}m^{2}-6\frac{\alpha^{4}}{a^{4}}\iota^{2}+2\frac{\alpha^{4}}{a^{2}}m\omega\iota^{2} \nonumber
\end{eqnarray}
\begin{eqnarray}
+6m^{2}\omega^{2}\alpha^{4}+\frac{\alpha^{4}}{a^{4}}\iota^{4}+10\frac{\alpha^{2}}{a^{2}}\left(E_{l,1}-m\right)C_{1}=0.
\label{kg20}
\end{eqnarray}

Equation (\ref{kg20}) is a fourth-order equation of $E_{l,1}$ from which one can find the expression of the ground state energy level of oscillator fields under a linear confining potential.

The ground state wave function is given by
\begin{eqnarray}
Q_{1,l}(z)=e^{\frac{a\,z\,(-a\,m\,\omega\,\alpha+\sqrt{2(E_{1,l}-m)\,C_{1}+a^{2}\,m^{2}\,\omega^{2}\,\alpha^{2}}}{2\alpha}} (c_0+c_1\,z).
\label{ee}
\end{eqnarray}

\subsection{Coulomb-Type Potential}

In this section, we consider another potential called Coulomb-type potential. This type of potential has widely been used to study various physical phenomena, such as the propagation of gravitational waves \cite{HA}, confinement of quark models \cite{gg1}, molecular models \cite{SMI}, and in quantum mechanics \cite{gg8,gg9,gg11,gg12,gg13,gg14}. Therefore, the Coulomb-type scalar (equal to vector potential) potential is given by

\begin{eqnarray}
V\left(z\right)=S\left(z\right)=\frac{C_{2}}{z}, 
\label{kg21}
\end{eqnarray}
where $C_2$ is a constant determines the strength of potential.

Thereby, substituting the Coulomb potential in Eq (\ref{kg11}), one will arrive the following equation 
\begin{eqnarray}
\frac{d^{2}Q\left(z\right)}{dz^{2}}+\left[a^{2}m\omega+\frac{1}{2z}+\frac{1}{\left(z-1\right)}\right]\frac{dQ\left(z\right)}{dz} \nonumber
\end{eqnarray}
\begin{eqnarray}
+\left[\frac{\iota^{2}-a^{2}b^{2}}{4z}+\frac{4a^{2}m\omega-\iota^{2}}{4\left(z-1\right)}+\frac{a^{2}\left[m-E\right]C_{2}}{2\alpha^{2}z^{2}}\right]Q\left(z\right)
\label{kg22}
\end{eqnarray}
The solution of above equation is given by
\begin{eqnarray}\label{kg6}
Q\left(z\right)=z^{\frac{\alpha-\sqrt{8\left(E-m\right)a^{2}C_{2}+\alpha^{2}}}{4\alpha}}\times \nonumber
\end{eqnarray}
\begin{eqnarray}
H_{c}\left(a^{2}m\omega,-\frac{\sqrt{8\left(E-m\right)a^{2}C_{2}+\alpha^{2}}}{2\alpha},0,\frac{a^{2}m\omega-a^{2}b^{2}}{4},\frac{a^{2}m\omega+1-\iota^{2}+a^{2}b^{2}}{4};z\right).
\label{kg23}
\end{eqnarray}

Substituting the power series solution given by 
\begin{eqnarray}
Q\left(z\right)=\sum_{j=0}^{\infty}c_{j}z^{j}
\end{eqnarray}
into the Eq. (\ref{kg22}), we obtain the following recurrence relation
\begin{eqnarray}
c_{j+2}= \nonumber
\end{eqnarray}
\begin{eqnarray}
\frac{\left\{ 4a^{2}m\omega j+5a^{2}m\omega-a^{2}b^{2}-a^{2}m\omega\frac{\sqrt{8\left(E-m\right)a^{2}C_{2}+\alpha^{2}}}{\alpha}\right\} }{4\left[\left(j+2\right)\left(j+1\right)+\left(\frac{2\alpha-\sqrt{8\left(E-m\right)a^{2}C_{2}+\alpha^{2}}}{2\alpha}\right)\left(j+2\right)\right]}c_{j} \nonumber
\end{eqnarray}
\begin{eqnarray}
+\left\{ 4j\left(j+1\right)+4\left(\frac{2\alpha-\sqrt{8\left(E-m\right)a^{2}C_{2}+\alpha^{2}}}{2\alpha}\right)\left(j+1\right)-4a^{2}m\omega\left(j+1\right)-\iota^{2}+a^{2}b^{2}\right. \nonumber
\end{eqnarray}
\begin{eqnarray}
\left.-a^{2}m\omega\left(\frac{\alpha-\sqrt{8\left(E-m\right)a^{2}C_{2}+\alpha^{2}}}{\alpha}\right)+\left(\frac{\alpha-\sqrt{8\left(E-m\right)a^{2}C_{2}+\alpha^{2}}}{\alpha}\right)+4\left(j+1\right)\right\}  \nonumber
\end{eqnarray}
\begin{eqnarray}
\times\frac{c_{j+1}}{4\left[\left(j+2\right)\left(j+1\right)+\left(\frac{2\alpha-\sqrt{8\left(E-m\right)a^{2}C_{2}+\alpha^{2}}}{2\alpha}\right)\left(j+2\right)\right]}.
\label{kg24}
\end{eqnarray}
With the coefficient
\begin{eqnarray}
c_{1}=\frac{a^{2}b^{2}-\iota^{2}-a^{2}m\omega\left(\frac{\alpha-\sqrt{8\left(E-m\right)a^{2}C_{2}+\alpha^{2}}}{\alpha}\right)+\left(\frac{\alpha-\sqrt{8\left(E-m\right)a^{2}C_{2}+\alpha^{2}}}{\alpha}\right)}{2\left(\frac{2\alpha-\sqrt{8\left(E-m\right)a^{2}C_{2}+\alpha^{2}}}{\alpha}\right)}c_{0}
\label{kg25}
\end{eqnarray}

let us consider $j=n-1$ when $c_{n+1}=0$ we have
\begin{eqnarray}
c_{n}= \nonumber
\end{eqnarray}
\begin{eqnarray}
\frac{\left\{ -4a^{2}m\omega\left(n-1\right)-5a^{2}m\omega+a^{2}b^{2}+a^{2}m\omega\frac{\sqrt{8\left(E-m\right)a^{2}C_{2}+\alpha^{2}}}{\alpha}\right\} c_{n-1}}{\left\{ 4n^{2}+4\left(\frac{2\alpha-\sqrt{8\left(E-m\right)a^{2}C_{2}+\alpha^{2}}}{2\alpha}\right)n-4a^{2}m\omega n-\iota^{2}+a^{2}b^{2}+\left(1-a^{2}m\omega\right)\left(\frac{\alpha-\sqrt{8\left(E-m\right)a^{2}C_{2}+\alpha^{2}}}{\alpha}\right)\right\} }.
\label{kg26}
\end{eqnarray}
Then, for $n=1$
\begin{eqnarray}
c_{1}=\frac{\left\{ -5a^{2}m\omega+a^{2}b^{2}+a^{2}m\omega\frac{\sqrt{8\left(E-m\right)a^{2}C_{2}+\alpha^{2}}}{\alpha}\right\} }{\left\{ 9-5a^{2}m\omega-\iota^{2}+a^{2}b^{2}+\left(a^{2}m\omega-3\right)\frac{\sqrt{8\left(E-m\right)a^{2}C_{2}+\alpha^{2}}}{\alpha}\right\} }c_{0}
\label{kg27}
\end{eqnarray}
By substituting Eq. (\ref{kg27}) into Eq. (\ref{kg25}), we have
\begin{eqnarray}
E_{l,1}^{4}+\left(\frac{6}{a^{2}}\alpha^{2}-10m\omega\alpha^{2}-\frac{2\iota^{2}}{a^{2}}\alpha^{2}-2m^{2}\right)E_{l,1}^{2} \nonumber
\end{eqnarray}
\begin{eqnarray}
+\left(2m\omega\alpha^{2}-\frac{2}{a^{2}}\alpha^{2}\right)\frac{\sqrt{8\left(E_{l,1}-m\right)a^{2}C_{2}+\alpha^{2}}}{\alpha}E_{l,1}^{2} \nonumber
\end{eqnarray}
\begin{eqnarray}
+\left(-10m^{2}\omega^{2}\alpha^{4}-2m^{3}\omega\alpha^{2}+\frac{2}{a^{2}}m^{2}\alpha^{2}+\frac{8m\omega}{a^{2}}\alpha^{4}-\frac{12}{a^{4}}\alpha^{4}-\frac{2m\omega\iota^{2}}{a^{2}}\alpha^{4}+\frac{4\iota^{2}}{a^{4}}\alpha^{4}\right)\frac{\sqrt{8\left(E_{l,1}-m\right)a^{2}C_{2}+\alpha^{2}}}{\alpha} \nonumber
\end{eqnarray}
\begin{eqnarray}
+m^{4}-\frac{6}{a^{2}}m^{2}\alpha^{2}+10m^{3}\omega\alpha^{2}+\frac{2\iota^{2}}{a^{2}}m^{2}\alpha^{2}-8\frac{m\omega}{a^{2}}\alpha^{4}+\frac{12}{a^{4}}\alpha^{4} \nonumber
\end{eqnarray}
\begin{eqnarray}
+\frac{10m\omega\iota^{2}}{a^{2}}\alpha^{4}+22m^{2}\omega^{2}\alpha^{4}+\frac{\iota^{4}}{a^{4}}\alpha^{4}-\frac{10\iota^{2}}{a^{4}}\alpha^{4} \nonumber
\end{eqnarray}
\begin{eqnarray}
+8\left(a^{2}m^{2}\omega^{2}\alpha^{2}-2m\omega\alpha^{2}+\frac{3}{a^{2}}\alpha^{2}\right)\left(E_{l,1}-m\right)C_{2}=0
\label{kg30}
\end{eqnarray}

Equation (\ref{kg30}) is a fourth-order equation of $E_{l,1}$ from which one can find the expression of the ground state energy level of oscillator fields under Coulomb-type interaction potential.

The ground state wave function is given by
\begin{eqnarray}
Q_{1,l}(z)=z^{\frac{\alpha-\sqrt{8\left(E_{1,l}-m\right)a^{2}C_{2}+\alpha^{2}}}{4\alpha}} (c_0+c_1\,z).
\label{ee2}
\end{eqnarray}

\subsection{Cornell-Type Potential}

Finally, here we choose another potential superposition of linear plus Coulomb-type potential. The Coulomb potential is responsible for the short ranges interactions and the linear potential leads to the confinement phenomena. This type of has been studied in the ground state of three quarks \cite{CA}, and in the relativistic and non-relativistic wave equations by several authors \cite{gg8,gg9,gg15,gg1,gg12,gg13,gg14}. Given this, let us consider this type of potential 

\begin{eqnarray}
V\left(z\right)=S\left(z\right)=C_{1}z+\frac{C_{2}}{z}
\end{eqnarray}
The Eq (11) can be rewrite as follows
\begin{eqnarray}
\frac{d^{2}Q\left(z\right)}{dz^{2}}+\left[a^{2}m\omega+\frac{1}{2z}+\frac{1}{\left(z-1\right)}\right]\frac{dQ\left(z\right)}{dz} \nonumber
\end{eqnarray}
\begin{eqnarray}
+\left[\frac{\iota^{2}-a^{2}b^{2}}{4z}+\frac{4a^{2}m\omega-\iota^{2}}{4\left(z-1\right)}+\frac{a^{2}C_{1}\left[m-E\right]}{2\alpha^{2}}+\frac{a^{2}C_{2}\left[m-E\right]}{2\alpha^{2}z^{2}}\right]Q\left(z\right)=0
\end{eqnarray}
we have
\begin{eqnarray}
Q\left(z\right)=z^{\frac{\alpha-\sqrt{8\left(E-m\right)a^{2}C_{2}+\alpha^{2}}}{4\alpha}}e^{\frac{az\left(-am\omega\alpha+\sqrt{2\left(E-m\right)C_{1}+a^{2}m^{2}\omega^{2}\alpha^{2}}\right)}{2\alpha}}\times \nonumber
\end{eqnarray}
\begin{eqnarray}
H_{c}\left(\frac{a\sqrt{2\left(E-m\right)C_{1}+a^{2}m^{2}\omega^{2}\alpha^{2}}}{\alpha},-\frac{\sqrt{8\left(E-m\right)a^{2}C_{2}+\alpha^{2}}}{2\alpha},0,\frac{a^{2}m\omega-a^{2}b^{2}}{4},\frac{a^{2}m\omega+1-\iota^{2}+a^{2}b^{2}}{4};z\right)
\end{eqnarray}

we write $Q\left(z\right)$ in Eq (22) 
\begin{eqnarray}
Q\left(z\right)=\sum_{j=0}^{\infty}c_{j}z^{j}
\end{eqnarray}
we obtain the recurrence relation
\begin{eqnarray}
c_{j+2}= \nonumber
\end{eqnarray}
\begin{eqnarray}
\frac{\left[\frac{4a\sqrt{2\left(E-m\right)C_{1}+a^{2}m^{2}\omega^{2}\alpha^{2}}}{\alpha}\left(j+1\right)-a^{2}b^{2}+a^{2}m\omega-\frac{\sqrt{8\left(E-m\right)a^{2}C_{2}+\alpha^{2}}}{\alpha}\frac{\left(a\sqrt{2\left(E-m\right)C_{1}+a^{2}m^{2}\omega^{2}\alpha^{2}}\right)}{\alpha}\right]}{4\left(j+2\right)\left[\left(j+2\right)-\frac{\sqrt{8\left(E-m\right)a^{2}C_{2}+\alpha^{2}}}{2\alpha}\right]}c_{j} \nonumber
\end{eqnarray}
\begin{eqnarray}
+\left\{ 4\left(j+1\right)\left(j+2\right)-\left(\frac{\sqrt{8\left(E-m\right)a^{2}C_{2}+\alpha^{2}}}{\alpha}+\frac{2a\sqrt{2\left(E-m\right)C_{1}+a^{2}m^{2}\omega^{2}\alpha^{2}}}{\alpha}\right)\left(2j+3\right)\right. \nonumber
\end{eqnarray}
\begin{eqnarray}
\left.-\iota^{2}+a^{2}b^{2}+a^{2}m\omega+1+\frac{\sqrt{8\left(E-m\right)a^{2}C_{2}+\alpha^{2}}}{\alpha}\frac{\left(a\sqrt{2\left(E-m\right)C_{1}+a^{2}m^{2}\omega^{2}\alpha^{2}}\right)}{\alpha}\right\}   \nonumber
\end{eqnarray}
\begin{eqnarray}
\times\frac{c_{j+1}}{4\left(j+2\right)\left[\left(j+2\right)-\frac{\sqrt{8\left(E-m\right)a^{2}C_{2}+\alpha^{2}}}{2\alpha}\right]}
\end{eqnarray}

with the coefficient
\begin{eqnarray}
c_{1}=\left\{ a^{2}b^{2}+a^{2}m\omega-\iota^{2}+1-\frac{\sqrt{8\left(E-m\right)a^{2}C_{2}+\alpha^{2}}}{\alpha}-\frac{2a\sqrt{2\left(E-m\right)C_{1}+a^{2}m^{2}\omega^{2}\alpha^{2}}}{\alpha}\right.\nonumber
\end{eqnarray}
\begin{eqnarray}
\left.+\frac{\sqrt{8\left(E-m\right)a^{2}C_{2}+\alpha^{2}}}{\alpha}\frac{\left(a\sqrt{2\left(E-m\right)C_{1}+a^{2}m^{2}\omega^{2}\alpha^{2}}\right)}{\alpha}\right\} \nonumber
\end{eqnarray}
\begin{eqnarray}
\times\frac{c_{0}}{4\left(1-\frac{\sqrt{8\left(E-m\right)a^{2}C_{2}+\alpha^{2}}}{2\alpha}\right)}
\end{eqnarray}
let us consider $j=n-1$ when $c_{n+1}=0$ we have
\begin{eqnarray}
c_{n}=\frac{A_{n}}{B_{n}}c_{n-1}
\end{eqnarray}
where
\begin{eqnarray}
A_{n}=a^{2}b^{2}-a^{2}m\omega-\frac{4a\sqrt{2\left(E-m\right)C_{1}+a^{2}m^{2}\omega^{2}\alpha^{2}}}{\alpha}n \nonumber
\end{eqnarray}
\begin{eqnarray}
\frac{\sqrt{8\left(E-m\right)a^{2}C_{2}+\alpha^{2}}}{\alpha}\frac{\left(a\sqrt{2\left(E-m\right)C_{1}+a^{2}m^{2}\omega^{2}\alpha^{2}}\right)}{\alpha}
\end{eqnarray}
and
\begin{eqnarray}
B_{n}=4n\left(n+1\right)-\left(\frac{\sqrt{8\left(E-m\right)a^{2}C_{2}+\alpha^{2}}}{\alpha}+\frac{2a\sqrt{2\left(E-m\right)C_{1}+a^{2}m^{2}\omega^{2}\alpha^{2}}}{\alpha}\right)\left(2n+1\right)-\iota^{2} \nonumber
\end{eqnarray}
\begin{eqnarray}
+a^{2}b^{2}+a^{2}m\omega+1+\frac{\sqrt{8\left(E-m\right)a^{2}C_{2}+\alpha^{2}}}{\alpha}\frac{\left(a\sqrt{2\left(E-m\right)C_{1}+a^{2}m^{2}\omega^{2}\alpha^{2}}\right)}{\alpha}
\end{eqnarray}
Then, for $n=1$
\begin{eqnarray}
c_{1}=\frac{A_{1}}{B_{1}}c_{0}
\end{eqnarray}
By substituting Eq. (38) into Eq. (34) we have
\begin{eqnarray}
E_{l,1}^{4}+\left(6\frac{\alpha^{2}}{a^{2}}-2\frac{\alpha^{2}}{a^{2}}\iota^{2}-2m^{2}-2m\omega\alpha^{2}\right)E_{l,1}^{2}\nonumber
\end{eqnarray}
\begin{eqnarray}
 \frac{8\alpha}{a}\sqrt{2\left(E_{l,1}-m\right)C_{1}+a^{2}m^{2}\omega^{2}\alpha^{2}}E_{l,1}^{2}-\frac{2\alpha}{a^{2}}\sqrt{8\left(E_{l,1}-m\right)a^{2}C_{2}+\alpha^{2}}E_{l,1}^{2}\nonumber
\end{eqnarray}
\begin{eqnarray}
 +\frac{2}{a}\sqrt{2\left(E_{l,1}-m\right)C_{1}+a^{2}m^{2}\omega^{2}\alpha^{2}}\sqrt{8\left(E_{l,1}-m\right)a^{2}C_{2}+\alpha^{2}}E_{l,1}^{2}\nonumber
\end{eqnarray}
\begin{eqnarray}
+26\frac{\alpha^{2}}{a^{2}}\left(E_{l,1}-m\right)C_{1}+\left(8a^{2}m^{2}\omega^{2}\alpha^{2}+24\frac{\alpha^{2}}{a^{2}}\right)\left(E_{l,1}-m\right)C_{2}+16\left(E_{l,1}-m\right)^{2}C_{1}C_{2} \nonumber
\end{eqnarray}
\begin{eqnarray}
+\left(\frac{2\alpha}{a^{2}}m^{2}-8m^{2}\omega^{2}\alpha^{3}-2\frac{\alpha^{3}}{a^{2}}m\omega+4\frac{\alpha^{3}}{a^{4}}\iota^{2}-12\frac{\alpha^{3}}{a^{4}}\right)\sqrt{8\left(E_{l,1}-m\right)a^{2}C_{2}+\alpha^{2}}\nonumber
\end{eqnarray}
\begin{eqnarray}
+\left(\frac{8\alpha}{a}m^{2}+8m\omega\frac{\alpha^{3}}{a}+8\frac{\alpha^{3}}{a^{3}}\iota^{2}-10\frac{\alpha^{3}}{a^{3}}\right)\sqrt{2\left(E_{l,1}-m\right)C_{1}+a^{2}m^{2}\omega^{2}\alpha^{2}}\nonumber
\end{eqnarray}
\begin{eqnarray}
+\left(10\frac{\alpha^{2}}{a^{3}}-2m\omega\frac{\alpha^{2}}{a}-2\frac{\alpha^{2}}{a^{3}}\iota^{2}-\frac{2}{a}m^{2}\right)\sqrt{2\left(E_{l,1}-m\right)C_{1}+a^{2}m^{2}\omega^{2}\alpha^{2}}\sqrt{8\left(E_{l,1}-m\right)a^{2}C_{2}+\alpha^{2}}\nonumber
\end{eqnarray}
\begin{eqnarray}
+2m\omega\frac{\alpha^{4}}{a^{2}}-10\iota^{2}\frac{\alpha^{4}}{a^{4}}+12\frac{\alpha^{4}}{a^{4}}+14m^{2}\omega^{2}\alpha^{4}+2m\omega\iota^{2}\frac{\alpha^{4}}{a^{2}}+\frac{\alpha^{4}}{a^{4}}\iota^{4}\nonumber
\end{eqnarray}
\begin{eqnarray}
+2m^{3}\omega\alpha^{2}+m^{4}-6m^{2}\frac{\alpha^{2}}{a^{2}}+2m^{2}\frac{\alpha^{2}}{a^{2}}\iota^{2}\nonumber
\end{eqnarray}
\begin{eqnarray}
-\frac{16\alpha}{a}\sqrt{2\left(E_{l,1}-m\right)C_{1}+a^{2}m^{2}\omega^{2}\alpha^{2}}\left(E_{l,1}-m\right)C_{2} \nonumber
\end{eqnarray}
\begin{eqnarray}
-\frac{16\alpha}{a^{2}}\left(E_{l,1}-m\right)C_{1}\sqrt{8\left(E_{l,1}-m\right)a^{2}C_{2}+\alpha^{2}}=0.
\label{kg42}
\end{eqnarray}

Equation (\ref{kg42}) is a fourth-order equation of $E_{l,1}$ from which one can find the expression of the ground state energy level of oscillator fields with a Cornell-type equal scalar and vector potentials.

The ground state wave function is given by
\begin{eqnarray}
Q_{1,l}(z)=z^{\frac{\alpha-\sqrt{8\left(E_{1,l}-m\right)a^{2}C_{2}+\alpha^{2}}}{4\alpha}}e^{\frac{ax\left(-am\omega\alpha+\sqrt{2\left(E-m\right)C_{1}+a^{2}m^{2}\omega^{2}\alpha^{2}}\right)}{2\alpha}}\,(c_0+c_1\,z).
\label{ee3}
\end{eqnarray}

\section{Conclusions}

In this work, we have investigated the generalized Klein-Gordon oscillator under the effect of equal scalar and vector potentials in a topologically charged Ellis-Bronnikov wormhole space-time. The basic idea was to study the behaviour of a relativistic scalar oscillator field with a position dependent mass in the wormhole space-time. The generalized KG-oscillator was introduced by repalcing the partial derivative $\partial_{\mu} \to \partial_{\mu}+M\,\omega\,X_{\mu}$ and inserting a scalar potential as the modification of the mass term $M^2 \to (M+S)^2$ in the Klein-Gordon wave equation. An interaction was considered using an electric charge and quadri-vector of the electromagnetic potential and the modified form of Klein-Gordon equation was obtained. The obtained equation was separated as a set of second order differential equations in terms of each spherical coordinate by using a suitable transformation and potential term. A trial solution was considered in the usual exponential form to solve the obtained differential equation. After applying a suitable transformation of coordinates the confluent Heun equation and the corresponding confluent Heun function was obtained. The obtained confluent Heun equation was solved by the Frobenius method as a power series expansion around the origin by considering a linear confining potential ({\bf Case A}) and obtained the ground state energy levels $E_{1,l}$ which satisfies the relation (\ref{kg20}) and the ground state wave function (\ref{ee}). Then, we have considered Coulomb-type scalar and vector potential ({\bf Case B}) and obtained the energy levels $E_{1,l}$ satisfying the relation (\ref{kg30}) and the ground state wave function (\ref{ee2}). Finally, we have chose a Cornell-type scalar and vector potential ({\bf Case C}) and obtained the energy levels $E_{1,l}$ satisfying the relation (\ref{kg42}) and the ground state wave function (\ref{ee3}). Following the similar procedure, one can obtain other energy levels $E_{2,l}, E_{3,l},...$ and the corresponding wave functions $\psi_{1,l}, \psi_{2,l},....$ for the radial modes $n=2,3,4,...$. Throughout the analysis, we have seen that the energy levels and the wave functions of oscillator fields are influenced by the topological defects of the wormhole space-time characterise by the parameter $\alpha$. Also, the parameter $a$ that represents the radius of the wormhole throat influences the eigenvalue solutions and modified the results.

\section*{Acknowledgement}
P.R. acknowledges the Inter University Centre for Astronomy and Astrophysics (IUCAA), Pune, India for granting visiting associate-ship.
 
\section*{Conflict of Interests}

There are no conflict of interests regarding publication of this paper.

\section*{Data Availability}

No new data are generated in this paper.

\section*{Funding Source}

No fund has received for this paper.


\begin{thebibliography}{99}

\bibitem{td1} A. Vilenkin and E. P. S. Shellard, {\tt Strings and Other Topological Defects}, Cambrigde University Press, Cambridge (1994).

\bibitem{td2} T. Vachaspati, {\tt Kinks and Domain Walls: An Introduction to Classical and Quantum Solitons}, Cambridge University Press, Cambridge (2006).

\bibitem{cs1} A. Vilenkin, Phys. Lett. B {\bf 133}, 177 (1983).

\bibitem{cs2} W. A. Hiscock, Phys. Rev. D {\bf 31}, 3288 (1985).

\bibitem{dl1} R. A. Puntigam and H. H. Soleng, Class. Quantum Grav. {\bf 14}, 1129 (1997).

\bibitem{dl2} V. B. Bezerra, J. Math. Phys. {\bf 38}, 2553 (1997).

\bibitem{gm1} M. Barriola and A. Vilenkin, Phys. Rev. Lett. {\bf 63}, 341 (1989).

\bibitem{gm2} A. L. Cavalcanti de Oliveira and E. R. Bezerra de Mello, Class. Quantum Gravity {\bf 23}, 5249 (2006).

\bibitem{gm3} A. Boumali and H. Aounallah, Adv. High Energy Phys. {\bf 2018}, 1031763 (2018).

\bibitem{gm4} H. Aounallah and A. Boumali, Phys. Part. Nucl. Lett. {\bf 16}, 195 (2019).

\bibitem{gm5} A. Boumali and H. Aounallah, Rev. Mex. Fis. {\bf 66}(2), 192 (2020).

\bibitem{kg1} E. A. F. Bragança, R. L. L. Vitória, H. Belich and E. R. Bezerra de Mello, Eur. Phys. J. C {\bf 80}, 206 (2020).

\bibitem{qho1} C. Furtado and F. Moraes, J. Phys. A Math. Gen. {\bf 33}, 5513 (2000).

\bibitem{qho2} R. L. L. Vitoria and H. Belich,  Phys. Scr. {\bf 94} 125301 (2019).

\bibitem{SR2} F. Ahmed, Sci. Rep {\bf 12}, 8794 (2022).


\bibitem{MM} M. de Montigny, H. Hassanabadi, J. Pinfold and S. Zare, Eur. Phys. J. Plus {\bf 136}, 788 (2021).

\bibitem{MM2} M. de Montigny, J. Pinfold, S. Zare and H. Hassanabadi, Eur. Phys. J. Plus {\bf 137}, 54 (2022).

\bibitem{imp} M. O. Katanaev and I. V. Volovich, Ann. Phys. (N.Y.) {\bf 216}, 1 (1992).

\bibitem{energy1} G. A. Marques and V. B. Bezerra, Class. Quantum Gravity {\bf 19}, 985 (2002).

\bibitem{nr1} E. R. Bezerra de Mello and C. Furtado, Phys. Rev. D {\bf 56}, 1345 (1997).

\bibitem{MP2} F. Ahmed, Mol. Phys. (2022), e2124935, DOI:10.1080/00268976.2022.2124935.

\bibitem{MP3} F. Ahmed, arXiv: 2209.13490 [quant-ph].

\bibitem{MP4} F. Ahmed, arXiv: 2210.04617 [hep-th].

\bibitem{sch1} N. A. Rao and B. A. Kagali, Phys. Scr. {\bf 77}, 015003 (2008).

\bibitem{dirac1} M. Moshinsky and A. Szczepaniak, J. Phys. A: Math. Gen. {\bf 22}, L817 (1989).

\bibitem{kkt} E. V. B. Leite, H. Belich and R. L. L. Vitoria, Braz. J. Phys. {\bf 50}, 744 (2020).

\bibitem{ads} B. Hamil and M. Merad, Eur. Phys. J. Plus {\bf 133}, 174 (2018).

\bibitem{st} R. L. L. Vitoria and K. Bakke, Eur. Phys. J. Plus {\bf 133}, 490 (2018).

\bibitem{css} A. Boumali and N. Messai, Can. J. Phys. {\bf 92}, 1490 (2014).

\bibitem{cp1} K. Bakke and C. Furtado, Ann. Phys. (NY) {\bf 355}, 48 (2015).

\bibitem{cp2} R. L. L. Vitoria and K. Bakke, Eur. Phys. J. Plus {\bf 131}, 36 (2016).

\bibitem{wh1} L. Flamm, {\it Physikalische Zeitschrift} {\bf 17}, 448 (1916).

\bibitem{wh2} A. Einstein and N. Rosen, Phys. Rev. {\bf 48}, 73 (1935).

\bibitem{wh3} M. S. Morris and K. S. Thorne, Amer. J. Phys. {\bf 56}, 395 (1988).

\bibitem{wh4} S. W. Hawking, Phys. Rev. D {\bf 46}, 603 (1992).

\bibitem{wh5} J. L. Friedmann, K. Schleich and D. N. Witt, Phys. Rev. Lett. {\bf 71}, 1486 (1993).

\bibitem{wh6} V. Frolov and I. D. Novikov, Phys. Rev. D {\bf 48}, 1607 (1993).

\bibitem{wh7} J. P. S. Lemos and F. S. N. Lobo, Phys. Rev. D {\bf 78}, 044030 (2008).

\bibitem{wh8} S.-W. Kim and H. Lee, Phys. Rev. D {\bf 63}, 064014 (2001).

\bibitem{wh9} F. Rahaman, P. K. F. Kuhfittig, M. Kalam, A. A. Usmani and S. Ray, Class. Quant. Grav. {\bf 28}, 155021 (2011).

\bibitem{wh10} M. S. Morris, K. S. Thorne and U. Yurtsever, Phys. Rev. Lett. {\bf 61}, 1446 (1988).

\bibitem{wh11}T. A. Roman, Phys. Rev. {\bf D 47}, 1370 (1993).

\bibitem{wh12} G. C. Samanta and N. Godani, Eur. Phys. J. C {\bf 79}, 623 (2019).

\bibitem{wh13} J. Sadeghi, M. Shokri, S. Noori Gashti, B. Pourhassan and P. Rudra, Int. J Mod. Phys. D {\bf 31} 03, 2250019 (2022)

\bibitem{eb1} H. G. Ellis, J. Math. Phys. {\bf 14}, 104 (1973).

\bibitem{eb2} K. A. Bronnikov, Acta Phys. Pol. B {\bf 4}, 251 (1973).

\bibitem{mg1} S. Nojiri and S. D. Odintsov, Int. J. Geom. Meth. Mod. Phys. {\bf 4}, 115 (2007).

\bibitem{de1} P. Brax, Rep. Prog. Phys. {\bf 81}, 1 (2018).

\bibitem{eibi1} M. Bañados and P. G. Ferreira, Phys. Rev. Lett. {\bf 105}, 011101 (2010).

\bibitem{eibi2} J. R. Nascimento, G. J. Olmo, A. Yu. Petrov, P. J. Porfirio and A. R. Soares, Phys. Rev. D {\bf 99}. 064053 (2019).

\bibitem{eibi3} R. D. Lambaga and H. S. Ramadhan, Eur. Phys. J. C {\bf 78}, 436 (2018).

\bibitem{eb3} H. Aounallah, A. R. Soares and R. L. L. Vitória, Eur. Phys. J. C {\bf 80}, 447 (2020).

\bibitem{eb4} A. R. Soares, R. L. L. Vitória and H. Aounallah, Eur. Phys. J. Plus {\bf 136}, 966 (2021).

\bibitem{gg1} C. L. Chrichfield, J. Math. Phys. {\bf 17}, 261 (1976).

\bibitem{gg2} I. C. Fonseca and K. Bakke, Proc. R. Soc. {\bf A 471}, 20150362 (2015).

\bibitem{gg6} E. Castro and P. Martin, J. Phys. A: Math. Gen. {\bf 33}, 5321 (2000).

\bibitem{gg3} R. L. L. Vitoria, C. Furtado and K. Bakke, Ann. Phys. {\bf 370}, 128 (2016).

\bibitem{gg4} M. Hosseinpour, H. Hassanabadi and F. M. Andrade, Eur. Phys. J. C {\bf 78}, 93 (2018).

\bibitem{gg5} H. Hassanabadi, M. Hosseinpour and M. de Montigny, Eur. Phys. J. Plus {\bf 132}, 541 (2017).

\bibitem{gg7} P. Sedaghatnia, H. Hassanabadi and F. Ahmed, Eur. Phys. J. C {\bf 79}, 541 (2019).

\bibitem{gg8} M. de Montigny, S. Zare and H. Hassanabadi, Gen. Relativ. Gravit. {\bf 50}, 47 (2018).

\bibitem{gg9} R. L. L. Vitória and K. Bakke, Eur. Phys. J. Plus {\bf 131}, 36 (2016).

\bibitem{gg15} E. R. F. Medeiros and E. R. B. de Mello, Eur. Phys. J. C {\bf 72}, 2051 (2012).

\bibitem{gg16} R. F. Ribeiro and K. Bakke, Ann. Phys. {\bf 385}, 36 (2017).

\bibitem{gg10} E. V. B. Leite, R. L. L. Vitória and H. Belich, Mod. Phys. Lett. {\bf A 34}, 1950319 (2019).

\bibitem{gg11} F. Ahmed, Eur. Phys. J. C {\bf 80}, 211 (2020).

\bibitem{gg12} E. V. B. Leite, H. Belich and R. L. L. Vitoria, Adv. High Energy Phys. {\bf 2019}, 6740360 (2019).

\bibitem{gg13} F. Ahmed, Sci. Rep. {\bf 11}, 1742 (2021).

\bibitem{gg14} E. V. B. Leite, H. Belich and R. L. L. Vitoria, Braz. J. Phys. {\bf 50}, 744 (2020).

\bibitem{HA} H. Asada and T. Futamase, Phys. Rev. {\bf D 56}, R6062 (1997).

\bibitem{SMI} S. M. Ikhdair, B. J. Falaye and M. Hamzavi, Ann. Phys. (N. Y.)  {\bf 353}, 282 (2015).

\bibitem{CA} C. Alexandrou, P. de Forcrand and O. Jahn, Nucl. Phys. B (Proceedings Supplements) {\bf 119}, 667 (2003).



\end{thebibliography}
\end{document}